# An Ontology-Based, Fully Probabilistic, Scalable Method for Human Activity Recognition

Pouya Foudeh[a,*] and Naomie Salim[a]
[a] *School of Computing, Universiti Teknologi Malaysia, Johor Bahru, Malaysia*
Emails: fpouya2@live.utm.my, naomie@utm.my

**Abstract.** Efficiency and scalability are obstacles that have not yet received a viable response from the human activity recognition research community. This paper proposes an activity recognition method. The knowledge model is in the form of ontology, the state-of-the-art in knowledge representation and reasoning. The ontology starts with probabilistic information about subjects' low-level activities and location and then is populated with the assertion axioms learned from data or defined by the user. Unlike methods that choose only the most probable candidate from sensor readings, the proposed method keeps multiple candidates with the known degree of confidence for each one and involves them in decision making. Using this method, the system is more flexible to deal with unreliable readings from sensors, and the final recognition rate is improved. Besides, to resolve the scalability problem, a system is designed and implemented to do reasoning and storing in a relational database management system. Numerical evaluations and conceptual benchmarking prove the proposed system feasibility.

## 1. Introduction

Human activity recognition (HAR) is one of the most applicable and yet challenging areas in context awareness. HAR is a multidisciplinary area of research, and it tackles several different issues. This paper proposes an ontological model for sensor-based human activity recognition that facilitates dealing with uncertainty. In some applications, HAR systems must be able to process a significant amount of data in a reasonable time. As the proposed system is highly scalable, it can be used for batch processing. A model for storing ontologies in relational databases is proposed in the form of tables that contain ontology's semantic material with accompanying probability values.

Sensor-based HAR is preferred for many applications since it is more pervasive, has fewer privacy issues, and requires less computational process and storage space compared to camera-based HAR. In sensor-based recognition, inertial measurement units (IMUs) and the indoor positioning system (IPS) track subjects' and environmental objects' movements and location. These sensors are basically analog devices that generate continuous signals. IMU sensors' data are processed with signal processing techniques; data-driven, classification methods convert them to posture (e.g., sitting or walking), and low-level, fine-grained activities (e.g., taking a coffee cup with the right hand). IPS signals are also used to calculate subjects' location in a room or building. This information is used to find out high-level activities, such as resting time, performed by the subject. Many algorithms have been presented for sensor-based HAR, from simple statistical models to deep learning [40].

In this paper, we proposed a HAR model. The main contributions of the current research are ability to deal with uncertainty, scalability and storing probabilistic ontological knowledge in ordinary relational databases. One of the challenges in HAR systems is uncertainty. The input information is essentially uncertain; sensors used in activity recognition are usually powered by unreliable batteries, data transmission is in a noisy wireless medium, and sensors might be displaced from their original position. Moreover, classification methods used for predicting low-level activities are not perfect. In short, there is no guarantee that whatever obtained from sensors' data is correct. However, the degree of belief, the probability of having correct information, is somehow calculable. When information such as low-level activities and location is available, one popular approach is to store them in a knowledge base system like an ontology system. Afterward, the knowledge base system does reasoning and infers high-level activities. In addition to probabilistic information obtained from sensors' data, known as ABox, the activity recognition knowledge base is also uncertain because the definition set, TBox, is also uncertain. For example, if Alice is certainly standing in place x in the kitchen taking a cup with her right hand and moving the chair with her left hand, she is probably in "tea-time" (80%), or she is in "cleaning time" (20%). This research models uncertainty with probabilistic representation and makes use of a probabilistic ontology to develop the proposed human activity recognition knowledge base. To the best of our knowledge, this is the first ontology-based activity recognition work with probabilistic observations from sensors.

Another aspect of HAR systems is their computing mode that can be real-time or batch processing. Real-time systems are for applications such as elderly monitoring and gaming, while batch processing is suitable for applications like employee monitoring, on parole criminal monitoring, and medical or praxeological studies on people's behavior. In real-time HAR systems, the processing time should be only less than or equal to the performing time and the window size should be small. However, batch processing recognition systems must be able to deal with a significant amount of data from several subjects, each performed in a long time span. Therefore, scalability is one of the key challenges in these systems. This paper proposes a probabilistic method for sensor-based human activity recognition to overcome both obstacles: uncertainty and scalability.

Storing data and reasoning about the knowledge base are difficulties of HAR batch processing. There are some knowledge management systems for storing and reasoning about conceptual knowledge bases. They are appropriate to use for a limited amount of data in research labs. On the other hand, relational database management systems (RDBMSs) are incredibly efficient, even though they are not designed to deal with complex knowledge structures. Decades of experience, billions of investments, and millions of active users have enabled RDBMSs to store and manage huge databases reliably and securely. For achieving scalability, this study designs a procedure to store activity recognition knowledge and do reasoning about them in a relational database. There are some research works on storing ontologies in databases. However, we didn't find any research on storing probabilistic ontologies on ordinary relational databases.

The organization of this paper is as follows. Section 2 reviews the relevant works on ontology-based activity recognition, probabilistic data, and knowledge bases and ontology storage in relational databases. Section 3 describes the applied dataset as the material and the applied methods for the probabilistic ontology model and reasoning. Section 4 is about experimental issues on software development and storing the knowledge base in a relational database. Section 5 reports the analysis, results, and comparisons, and Section 6 contains the conclusion and possibilities for future works.

## 2. Background

### 2.1. Ontology-based HAR

Recent works on sensor-based high-level human activity recognition fall into two categories: 1) data-driven recognition that approaches the problem from pattern recognition or machine learning view point and 2) knowledge-driven recognition that applies logical reasoning and, in particular, knowledge-based systems. The probabilistic Markov model is frequently used in data-driven approaches. [5] used this model for predicting users' situation, while providing low-level activities to the system, and [29] proposed a location-based activity recognition using the Markov model. In some recent studies, including [30] and [19], the Markov model and knowledge bases have been applied together, called the hybrid approach. Logical reasoning has been used to recognize and predict human activities from the long past [24]. Only It is regarded a huge progress after modern knowledge representation models are developed. Nowadays, ontologies are one of the best tools for activity recognition purposes [47]. Despite numerous studies on HAR, RapidHARe [12] is the only method we know which deals with processing time in this area.

Chen and Nugent [10] proposed an ontology-based approach that was one of the first integrated frameworks for activity recognition based on a conceptual essence. They did not use any dataset and presented the framework with a sample ontology. However, in an extension of their work [11], later they presented an activity recognition method based on the previously presented ontological model. The method is designed to confront a cold start, a common problem in data-driven activity recognition, when there are no or few labeled data for learning the system at the beginning point. At the starting point, there is an ontology developed according to human knowledge. Using the ontology, the system labels some activities, and then, new activities are discovered using labeled information and data-driven techniques. The labeled information is used to discover more activities via data-driven learning, and discovered activities are used to populate the ontology. More information is labeled by running all phases more than once. As a result, the ontology becomes more completed. In this method, the ontology and the information are deterministic, although the machine learning process is probabilistic.

Riboni and Bettini [45] proposed an ontology web language (OWL) 2 model for human activities. They also developed another hybrid, statistical, and ontological activity recognition method [44]. It was published simultaneously with Chen and Nugent's method, and both claimed their models to be the first in the field. This method obtains low-level activities using data-driven techniques from the body and environmental sensors and gathers the subject's location from GPS for outdoor and RFID for indoor. They used a novel technique, called historical variant, to temporally optimizing obtained low-level activities. Afterward, they utilized ontological reasoning, only for locations of objects, to predict high-level activities. Input and output data and the reasoning process were not probabilistic, and the most probable state was always chosen as the deterministic answer. Palumbo et al. [39] proposed a HAR process that is implemented within the reservoir computing paradigm. Ni et al. [37] proposed a HAR model consists a network of ontologies that aims smart homes especially for the elderly in the healthcare domain.

The probabilistic essence of human activity recognition calls for probabilistic logic and reasoning [42]. Nevertheless, early attempts preferred to avoid probabilistic logic because of its difficulties [3]. The most recent research attempts to avoid probabilistic ontologies because there is no established model and standard for that (see section 2.2). There are a few studies about adopting ontologies and probabilistic models and reasoning in activity recognition. Yamada et al. [57] was a preliminary research on using ontology in activity recognition that applied probabilistic modeling. Environmental objects are equipped with RFID tag and RFID reader sensors installed in the activity area. They track objects' location. However, because of the overlap between activity spaces and the RFID system unreliability, the process is probabilistic.

Helaoui et al. [23] is more similar to the current research, not only for using ontology and a probabilistic model but also for utilizing the Opportunity dataset, which is the earlier version of the dataset used in the current study. The dataset is about daily morning activities; the subjects' bodies and environmental objects are covered with different kinds of sensors, and RFID tags are installed in their gloves to track the subjects' location. In a newer version of the dataset used in this research, the state-of-the-art IPS system is utilized for indoor tracking. In addition to sensor data, the dataset contains labels for postures and low- (right and left hand interaction with objects and hand movements) and high-level activities. Dataset developers made annotations via video checking. Additionally, this particular research annotates two medium-level activities: simple activities and manipulative gestures. The probabilistic ontological reasoning process is as follows: information about location and posture is ignored, and hands' interaction with objects and hands' movements are combined and named atomic gestures (Level 4). A set of related and sequential atomic gestures characterize a manipulative gesture (Level 3); for example, "opening a drawer" and "fetching a knife" is "taking the knife". In the same way, a set of manipulative gestures characterize a simple activity (Level 2), and a set of simple activities characterize a high-level (complex) activity (Level 1). Level 4 labels are available in the dataset, meaning that regardless of errors, they are somehow obtained from sensor data. For levels 3 to 1, the state of each level is deducted from lower level using ontological reasoning. The axioms, Tbox, of ontology are manually developed and weighed as confidence property, the probability of working of the assertion axiom. The manually annotated labels for levels 1 to 3 are used to evaluate the method. There are two points to note about this research. First, the input and output of each step are not probabilistic. In other words, the reasoner with probabilistic rules receives deterministic values from the lower level and only passes the most probable value to the higher level (similar to the aforementioned works). Second, model reusability is not achievable in this method. For each individual with a specific lifestyle, the system needs a special set of simple activities and ontology assertion axioms.

One of very few works on activity recognition with uncertain observations is Roy et al. [49]. Similar to the previous research, they designed a multilevel reasoning model. However, the model is not ontology-based, and their approach is possibilistic, and not probabilistic, meaning that each attribute's value is characterized by a degree of certainty. In other words, each level (including the lowest level: sensors' reading) passes one or more values to the higher level or no value at all for total ignorance. The model is designed for recognizing only one specific activity: taking the medication by the patient. Therefore, unlike general activity recognition systems, the design of activities for different levels and the reasoning system for them is feasible. In this research, activities and events are modeled based on possibilistic networks.

The probability theory is not the only approach for measuring and dealing with uncertainty, and other approaches include belief functions, possibility theory (based on fuzzy logic), and plausibility theory used to model problems with an uncertainty of different nature [21]. Besides the earlier mentioned method, there are some studies on activity recognition that used non-probabilistic models to confront challenges made by an uncertain essence of an activity recognition task. For example, Rodríguez et al. [48] utilized fuzzy ontologies, Roy et al. [49] proposed a possibilistic reasoning method, and Noor et al. [38] developed an ontological reasoning process with belief functions (Dempster–Shafer) theory. All these studies are about knowledge-driven human activity recognition.

Ontologies, in the original style and with OWL language, do not support temporal reasoning. Without temporal reasoning, the knowledge-driven system will recognize activity instances as independent parts of information. In this approach, the efficiency decreases [46]. There are some proposals to add time information in RDF language [20]. For filling this gap, Meditskos et al. [35] proposed a framework for combining OWL and SPARQL to be used for activity recognition. SPARQL is a query language suitable for querying knowledge bases and capable of handling temporal relations. In [34], they extend the framework and combined it with a semantic activity model, and finally in [33], they presented a similar framework with SPARQL and implemented an activity recognition system according to it in the field of healthcare to monitor people with dementia. SQL is another query language used in the current research. Similar to SPARQL, it can deal with temporal information. However, relational databases' querying systems, including SQL language, do not have functions to support semantic knowledge querying directly.

### 2.2. Probabilistic Data and Knowledge Bases

There are several situations that must be dealt with probabilistic data. For example, information retrieval systems from textual corpus produce probabilistic data. Because of uncertainty on knowing the fact in the text or imperfectness of information retrieval methods, there are some discovered relationships without one hundred percent confidence. Such a data needs to be stored and queried efficiently. Therefore, the database community proposed probabilistic databases to answer this demand.

Depending on the probabilistic relational database model, the probability might appear in some properties, the tuple level, or a group of tuples known as block level. Whatever model is used, the amount of probability is stored as an extra property in the database. Query processing is not as easy as storage. Unlike conventional databases that work with one world, probabilistic databases deal with a numerous number of possible worlds. In this condition, it is impossible to compute some queries since they are hard for #P, the counting version of NP nondeterministic polynomial time, especially in a system that is supposed to be scalable, i.e. it cannot limit the amount of data. Probabilistic relational database management systems (PR-DBMSs) must determine whether they can evaluate the query before attempting to execute it. In addition to probabilistic databases, probabilistic data can be stored and reasoned via a graph in the form of Bayesian networks or Markov networks. In this case, the complexity is the tree-width, meaning that there is a non-computable query. On the other hand, the data model is more complicated, and data modification in a static graph is not as easy as adding some tuples to the database [52].

Ontologies have been used as a tool for knowledge organization. As a conceptual model, (non-probabilistic) ontologies are visually modeled in the form of graph regardless of how they are actually implemented and stored. Indeed, knowledge graph is another name for ontology. OWL is a language for representing ontologies. There are a few frameworks proposed for probabilistic ontologies. The most cited framework among them is PR-OWL [14][6]. It is a Bayesian framework probabilistic ontology that is based on multi-entity Bayesian networks [27]. Unlike OWL, PR-OWL is not endorsed by W3C as a standard language. It suffers from some problems: most specifically, it is not fully compatible with OWL. Besides, regular ontologies and OWL (not PR-OWL) are supported by well-established software, and Protégé is the most famous one. Nevertheless, PR-OWL-based ontologies have some applications, for example, in the automation of procurement fraud detection in Brazil [7] or duplicate publication and plagiarism detection [17].

In ontology-based human activity recognition, there are some probabilistic attempts. However, only the reasoning engine, and not input/output information, is probabilistic. Scalability and accuracy are priorities of the current research. The system is modelled in the form of a fully probabilistic ontology to reach a high level of accuracy. For scalability, highly efficient commercial RDBMSs are used for data storage and processing. PR-OWL tools or the research-based PR-DBMSs could help us implement this model more easily. However, they are not sufficiently reliable and efficient to manage such a big data. Even though the proposed system is modeled and implemented without adopting PR-OWL and PR-databases, the current work is partly inspired by both.

*2.3. Ontology Storage in Relational Databases*

There are similarities and dissimilarities between ontologies and databases, which can be categorized as information conceptualization, data representation (tuples vs. instances), data modeling, and, in practice, efficiency. In this regard, a new concept appeared: ontologies based on databases. The concept means using a relational data model to store data represented in an ontology [32]. Initial works in this field focused on proposing algorithms for transforming information and modeling from an ontology to a relational schema [18][54][55] and satisfying rules of relational databases such as primary and foreign key and data types [2]. Further studies confront other problems in this area, including query processing and optimization [22][1] and few works on inference [4]. Pipitone et al. [43] is a model for estimating the most likely alignment between an OWL ontology and an entity relation diagram and relies on a hidden Markov model, not Bayesian interpretation of probability. There is a research on probabilistic ontologies and relational databases [53]. However, the idea is based on modification of relational algebra, similar with probabilistic databases. In other words, it is no applicable to ordinary relational databases.

The Opportunity dataset has been used in some studies, enabling comparison of the approach's recognition rate between this study and others. All of them used manual labels of low-level activities, and because there was no label for the location, they just ignored it. On the other hand, this research uses probabilistic information computed from sensors' data. [31] applied a different classifier on the dataset to find out which classifiers had the best performance. [50] applied two modern methods for activity recognition: neutrosophic [51] lattice and fuzzy lattice.

This research approach is to store probabilistic ontologies in regular relational databases. There has been no literature on this issue so far, and the storing model is slightly different from methods for storing regular ontologies in regular databases. The main difference is that all tuples in Tbox and Abox have an extra probability property. Moreover, database constraint rules, including the primary key, are different. Although, the inference process for probabilistic ontologies is entirely different from the reasoning process about non-probabilistic ontologies. Thus, the proposed model is a single purpose system for activity recognition. Running general queries on probabilistic ontologies in databases is out of the scope of this research.

## 3. Material and Methods

In this section, the chosen dataset for this research is discussed and the probabilistic ontological method for HAR information is explained. This information initially includes low-level activities and subjects' locations. However, after the inference process, it will be populated to an ontology that contains high-level activities. The proposed data smoothing methods increase the accuracy of the results.

*3.1. Applied Dataset*

"Activity and context recognition with opportunistic sensor configuration", or Opportunity in short, is an EU project for developing a HAR dataset [9]. This dataset is a collection of data from several sensors, including accelerometers, gyroscopes, magnetic sensors, DIP switches, and an indoor tracking system, installed on subjects' bodies and environmental objects. The dataset is collected from four subjects, each performing daily morning activities five times, each time about half an hour, and one drill that is repetitions of predefined acts. All activities are video recorded and using these videos, the dataset is manually labeled. In addition to raw sensor values and signals, the dataset contains labels for: 1) postures (e.g., sitting), 2), 3) right- and left-hand movements (e.g., unlocking), 4), 5) right and left hand-object interactions (e.g., dishwashing), 6) gestures (e.g., opening a dishwasher), and X) high-level activities (e.g., sandwich time). The proposed system aims to predict X.

In the previous work [16], the Opportunity dataset was used to generate a new dataset with probabilistic predictions. Fast signal processing methods, suitable for batch processing, are applied to compute proper features from sensor signals. For each subject, the first three activities and the drill are used as a training set, and the system gives some probabilistic predictions for labels 1 to 6. High-level activities are more complicated to be efficiently predicted by a data-driven method. Figure 1 reports the probability of detection in different number of candidates. This figure justifies adopting probabilistic predictions. While the relying on the first choice is not very efficient, the probability of having the right answer in the first three choices, in the 18 choices of gestures, is about 0.9. In addition to predicting these six labels for the test set, the research proposes methods for calculating two more properties that are not labeled in the dataset. 7) Compass: it refers to the direction of subjects' bodies (non-probabilistic). 8) Location: the performance room is divided into 64 rectangle regions, 8 × 8, and the system must detect which rectangle is the current location of the subject. Because of the noisy nature of indoor tracking systems, the system makes a probabilistic guess, even though four independent tags locate the subject independently.

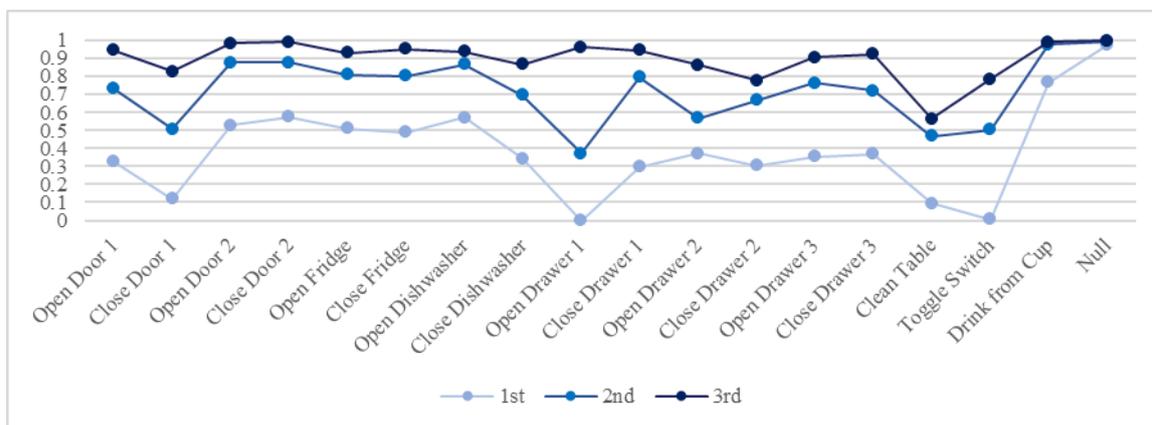

Fig. 1. The probability of detection in the first, first two, and first three choices of the gestures.

The current research will not use all calculated probabilistic properties obtained in the previous research. Since posture recognition accuracy was very high, it is used as a non-probabilistic property, and only the most probable choice is used. In practice, we observe that hand movements do not increase the recognition rate of high-level activities. It makes sense in a way that when the subject interacts with a cup, they are in the middle of breakfast or are involved in a cleaning activity, no matter if they reach or release the cup.

All the information, belongs to four subjects, is imported from the dataset to the ontology in the form of RDF, resource description framework, semantic triples. There are three instances per second; each instance has an instance number unique in the whole dataset (all instances have a fixed duration of 0.33 seconds). There are three properties of elemental information of an instance. 1) ID: is a unique instance number in the whole dataset. It is the first item of all triplets. 2) Serial: is a number referring to the activity number. In this case, the first digit presents the subject number (1 to 5), and the second digit shows the sequence number of the activity, with 1, 2, and 3 being daily morning activities, 9 being a drill, all used for training, and 4 and 5 being daily morning activities aiming to recognize their high-level activities. 3) Instance time: it refers to the beginning time of instance in seconds.

The predicted and calculated labels are merged to generate two atomic codes describing subjects' situation. 1) Location, Angle, and Posture (LAP), a four-digit code: the first two digits indicate the 2-dimension location of the subject in the room, the third digit is for the subject's angle, and the fourth digit indicates the subject's posture. The only probabilistic item in these groups is location. Therefore, the probability of location is the probability of instance. For example, for "11067" id, the LAP code is "6402" with a probability of "0.2". It means that during the instance number 11067, the subject location is in x=6, y=4 rectangle a chessboard-like room; the subject's face is between 0 ° (north pole) and 59.9 °, and her posture is walking. 2) Both hands object interactions (BHO): During each instance, the subject's both hands might interact with two different objects. Also, it is possible that one or both hands are idle. For example, for "11067" id, the BHO code is "2000" with a probability of "0.746", showing that during the instance number 11067, the subject's right hand interacts with the fridge and her left hand is idle. Right- and left-hand interactions are separate activities, and each hand has its own probability. Therefore, the probability of both interactions occurring is the product of them. In this example, the probability of being idle for the left-hand is 0.91 and being in interaction with the fridge for the right-hand is 0.82. Probabilities less than or equal to 0.01 are not used in the calculation. In addition to the mentioned classes, Null, with 0 code, is used when the class has no value; for example, the location is not calculable, or the posture is between sitting and standing.

*3.2. Ontology Model, Population and Constraints*

In the presented model, the RDF triples constitute the foundation of the primary ontology. The ontology is drawn using Eddy [28]. Since the ontology will be implemented in a relational database and we are not going to code it in OWL, Eddy is a suitable graphical tool for our purpose. It also guarantees the syntactic correctness of the design. As shown in Figure 2, each instance has a unique ID and two other attributes: serial and Instance time plus LAP (location, angle, and posture) and BHO (both hands-objects interactions) made by combining LHO, the left hand, RHO, and the right hand. High-level activities are empty at the starting point for under investigation (testing) instances and must be assigned during the reasoning process. For available training instances, BHO is deterministic and high-level activities are known; both are manually labeled. In the primary ontology, each instance (triple) represents 0.33 seconds of an activity and is independent from other instances. After starting the process, the ontology is extended. Some triples are replaced with more exact ones, assertion axioms are added to the ontology, high-level activities are formed gradually, and, finally, the secondary ontology containing high-level daily activities is developed. In other words, the semi-automatic ontology population process, adding new instances of concepts to the ontology [41], is performed by step by step data fusion. Integrating data starts from fine-grained independent instances and ends with an ontology of coarse-grained activities.

Ontologies are modeling tools for the semantic web. AAA Slogan (Anyone can say Anything about Any topic) and Open World Assumption (some statements have not been said yet) are regular conditions in the web data environment. For example, both statements are acceptable: "Taj Mahal is in China" and "Taj Mahal is in Afghanistan", although, in fact, "Taj Mahal is in India". In other words, inconsistency is accepted, and there is no constraint for the restriction of such data. In the current probabilistic ontology, inconsistency is also acceptable, although it is limited by probability rules. Inconsistent statements are accepted only when the summation of their probabilities is less than or equal to one. The open word is limited to one minus the summation of known probabilities. For example: "Taj Mahal is in China" with a probability of 0.3 and "Taj Mahal is in Afghanistan" with a probability of 0.2 are acceptable because (0.3+0.2) is ≤ 1 and the probability of being in anywhere, the open world space, is 1 – (0.2+0.3). The implemented ontology must enforce this constraint. In the current ontology, for each unique instance, the summation of LAP or BHO probabilities must be less than or equal to one.

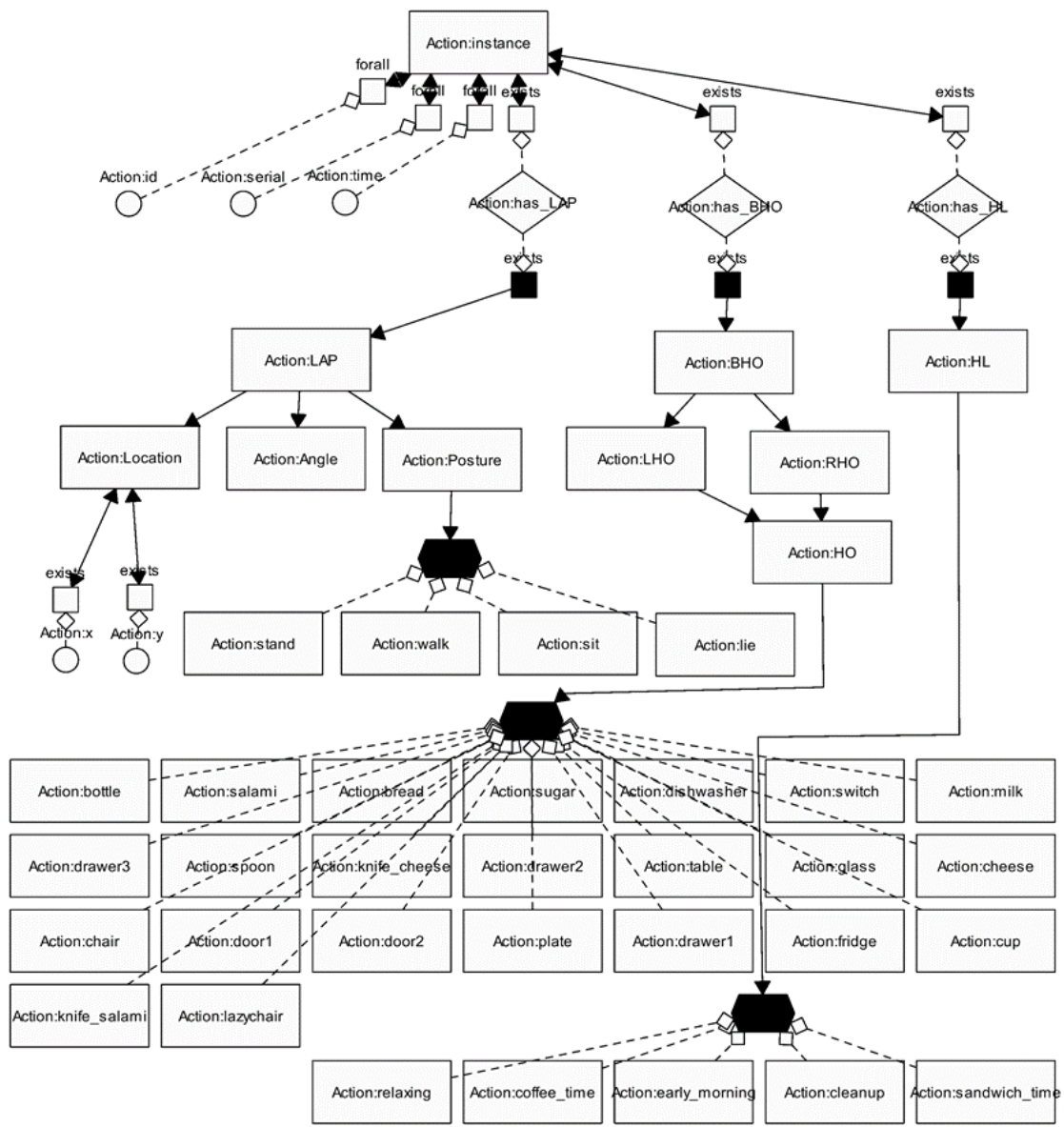

Fig. 2. The primary ontology formed from triples coming from signal processing.

*3.3. Probabilistic Data Smoothing*

All instances are computed only from sensors' reading during that particular instance. In other words, each instance is independent of others. This is not very realistic because the instance time is much shorter than the duration of an activity, and it is more likely that consecutive instances have the same value. For example, <walk, walk, lie, walk, walk> are five consecutive instances. It appears that the fourth instance is not correct and must be replaced. In this process, called data smoothing, data not in a normal pattern (outliers) are altered. The moving average is one of the most common data smoothing algorithms; the value of each point is replaced with the mean value of an interval with the point in its center or in real time systems at the end of it. For non-numeric data, the mode of data at the interval can be used instead of the mean value. However, this algorithm is for deterministic data, while this research deals with probabilistic data.

There is limited literature on probabilistic data smoothing. Merigó et al. [36] research is based on expertons theory and fuzzy logic. Hiemstra et al. [25] developed a probability smoothing technique for language modelling, text information retrieval. Kim et al. [26] is another research that is for target tracking in a surveillance space. These methods cannot be used for the purpose of this research. In addition to fundamental incompatibilities, it is for two states (win and lose), while the activity recognition system of this research must deal with multiple states in both LAP and BHO. As the current system is developed for batch processing, access to future instances is feasible. The probabilistic version of statistical mode to be used in the interval is introduced.

As a new operator for aggregating information [56], Probabilistic Mode (PM) is defined.

Let $\{A_1, A_2, ...., A_n\}$ be a collection of probabilistic arguments and $D = \{v_1, v_2, ...., v_m\}$ a countable domain. Each $A_i$ is a set of pairs; the first part is the value, from domain $D$, and the second part is its probability. For example, $D = \{sitting, standing, walking, lying\}$ and $A_3 = \{<sitting,0.6>, <standing,0.3>\}$. The probability of what is not mentioned in this set, walking and lying in this example, is 0, in known probability areas. When the summation of probabilities is less than 1, there is an open world space, 0.1 in this example, that belongs to all the domain members. $PR_{ai\_j}$ is the probability of $v_j$ in $A_i$. In this example, $PR_{a3\_1} = 0.6$, $PR_{a3\_2} = 0.3$, $PR_{a3\_3} = 0$, and $PR_{a3\_4} = 0$.

In the following formula, n is the size of the interval, m is the cardinality of the domain, B is a temporary variable, containing a set of pairs same as each $A_i$, and f is the Probabilistic Mode (PM) operator.

if B = f ({$A_1, A_2, ...., A_n$}) then
    for j = 1 to m  $\{PR_{B\_j} = (\sum_{i=1}^{n} PR_{Ai\_j})/n\}$    (1)

The regular mode function, non-probabilistic, can be defined according to the probability theory: mode is the value in the set that is most probable to be sampled. The defined probabilistic mode function is compatible with the regular mode; with deterministic inputs, each sample has one element with the probability of 1 and others are 0, and it will work same as the regular mode function if the most probable element of the result is taken. However, the probabilistic mode returns a probabilistic set, in the same format with arguments of the function input. Loosely speaking, the probabilistic mode (PM) calculates the average probability of each domain value. The PM function deals with known probabilities and does not consider the open world space.

The PM function is used for data smoothing of probabilistic properties of an activity instance. For BHO, the interval size is 5, and the current instance is replaced with the PM of 11 instances: five instances before, five instances after, and the current instances. For LAP, the interval size is 3. The interval length is lower for instances located at the beginning or end of an activity serial. After the data smoothing process, there is still a probabilistic set of values for instances. However, the probabilities are updated, and outlier data are inconspicuous.

*3.4. Obtaining and Defining the Assertion Axioms*

A set of assertion axioms is needed to attain the reasoning process. The premises, left side, is what is already known from the applied dataset, and the conclusion, right side, is what is going to be discovered, which is either true or false. The following simple example is an assertion that leads to the activity from the location:

(Bob is in the bed) → (Bob is in a resting time)

In a real world situation, the axioms of HAR systems are probabilistic. If Bob is in bed, he is in a resting time with a probability of 0.80 and he is in a reading time with a probability of 0.2. Axioms like this can be developed in two different ways. An expert user, a person who knows Bob's habits in this example, may define them. They also can be obtained from a training dataset. In this simple example, Bob's bedroom area can be video recorded for some time, e.g. 3 days, and then, his activities in the bed are statistically investigated.

In the current research, for each instance, there are two types of premise data: BHO (both hands-objects interactions) and LAP (location, angle, and posture). It is necessary to have a set of assertion axioms that lead from possible instances of BHO or LAP to high-level activities. Both methods, defining by the user and obtaining from the training data, are provided in the proposed system. The software system for defining the axioms will be discussed in Section 4.

Assertion axioms can also be automatically obtained, which is known as ontology learning. In this case, some manually labeled data are needed. For BHO, we have everything we need; manually labeled properties for high- and low-level activities, including interaction with objects with right and left hands, are available in the Opportunity dataset. For LAP, there are labels for high-level activities; however, there is no label for locations and angles. We suppose what is calculated for angles is correct, and the most probable location and posture is assumed to be genuine. The naïve Bayes classification algorithm is applied. Axioms are calculated according to the Kolmogorov definition:

$$P(A|B) = \frac{P(A \cap B)}{P(B)} \quad (2)$$

For example, let's take a training dataset with 1000 instances. We are going to calculate the probability of being in "sandwich time" if "right hand is in interaction with bread".

$$P(sandwich\_time|RH\_bread) = \frac{\frac{\text{Number of instances in "sandwich time and right hand interacts with bread"}}{1000}}{\frac{\text{Number of instances in "right hand interacts with bread"}}{1000}} \quad (3)$$

For all available, not all possible, BHO and LAP codes, the probability of being in each high-level activity can be calculated. There are two options to determine probabilities. One is manual defining that needs a user as an expert who knows all activity environment details. The other is automatic obtaining that requires a large amount of training data to collect a fairly accurate set of assertion axioms. Both methods have their own drawbacks. In practice, the combination of both methods yields the best performance. First, axioms are obtained using the automatic learning process. Then, a user can visually check them and do some modification. For example, there is a seat in the room that subjects may use it while eating. However, in the applied training dataset, accidentally no one has used it. In this case, the expert user can edit learned axioms to add this particular space to the eating area. There is a similar research, newNECTAR [13], that introduces a collaborative active learning process to refine the correlations in a HAR system. It obtains personalized temporal patterns of sensor events from the feedback.

*3.5. Inference Process*

Up to the current stage, there are some assertion axioms and some instances. Each instance has three or fewer BHO and three or fewer LAP states accompanying their probability values. For each BHO or LAP state, there is an assertion axiom, connecting it to the five probable high-level activities: relaxing, coffee time, early morning, cleanup, and sandwich time. After multiplexing probabilities of axioms and states and then the Cartesian product of BHO and LAP states, there are nine or fewer items for each instance:

$<PR_{BHO\_relax}$, $PR_{BHO\_coffee}$, $PR_{BHO\_morning}$, $PR_{BHO\_clean}$, $PR_{BHO\_sandwich}>,<PR_{LAP\_relax}$, $PR_{LAP\_coffee}$, $PR_{LAP\_morning}$, $PR_{LAP\_clean}$, $PR_{LAP\_sandwich}>$

For each item, there are two sets of probabilities of all high-level states: one obtained from BHO and another obtained from LAP. Since BHO and LAP are from different and independent processes, the combination of probabilities, e.g. for relaxing, can be calculated according to:

$$PR_{relax} = 1 - ((1 - PR_{BHO\_relax}) * (1 - PR_{LAP\_relax})) \quad (4)$$

Even though the above formula is theoretically intact, in practice, some modifications improve the prediction accuracy of high-level activities. In fact, BHO and LAP do not have equal affection on a high-level activity, and it should be manually adjusted. For each high-level activity, the weighting coefficients, m and n, are defined between 0 and 1, and the formula is rewritten as follows:

$$PR_{relax} = 1 - ((1 - m * PR_{BHO\_relax}) * (1 - n * PR_{LAP\_relax})) \quad (5)$$

According to the experiment of this research, the reported coefficients in Table 1 leads to relatively better results.

Table 1.
Coefficients for high level activities

|   | Relax | Coffee | Morning | Clean | Sandwich |
|---|---|---|---|---|---|
| m | 0.7 | 0.5 | 0.5 | 0.7 | 0.5 |
| n | 0.7 | 0.7 | 0.4 | 0.9 | 0.6 |

At this point, for each instance there are nine or fewer items, each having a probability for each high-level activity. Despite the whole process that was fully probabilistic, the final step should determine a deterministic suggestion for the high-level activity that the subject is performing during the instance. For each high-level activity, the highest probability value from all items is chosen, and then, the most probable high-level activity is the final candidate for the instance. Table 2 presents the actual calculation of the probability of one particular instance (14940) using LAP, BHO, and a total combination of both.

After obtaining a high-level activity for each instance, we can work on a group of instances to develop the final ontology. The final ontology has the same elemental information, ID, serial, and starting time, as well as a high-level activity label. There is no need for fine-grained labels, BHO and PAL. In this ontology, the duration of instances varies; therefore, an extra property, length, is needed to indicate the duration of the instance in seconds.

It is very likely that the high-level activity of a particular instance is the same as instances before and after it. This situation is similar to BHO and PAL properties of instances; however, in this case, data smoothing methods cannot be used. A low-level hand activity or physical position of the subject usually takes a few seconds, while there are three instances in one second. Although, a high-level activity usually takes several minutes, i.e. hundreds of instances.

Here is the algorithm for developing the final ontology and improving its prediction accuracy of high-level activities by eliminating some potential prediction errors. It is named three-step elimination. It removes some blocks of the predicted activities in three steps.

```
new_ontology = old_ontology
rearrange (new_ontology)
delete instances with length <15
rearrange (new_ontology)
delete instances with length <35
rearrange (new_ontology)
delete instances with length <55
rearrange (new_ontology)

define rearrenge(an ontology)
FOR current_instant From last_instant TO one_after_first_instant DO
    IF HL_activity (current_instant) = HL_activity (previous_current_instant)
        Delete (current_instant)
FOR current_instant From first_instant TO last_instant DO
    length (current_instant) = id(current_instant)- id(next_current_instant)
```

After rearranging the ontology, all consecutive instances with the same high-level activity are replaced with only one instance. In other words, a large ontology with plenty of instances is converted to an ontology with a few instances. It will be even smaller and also more accurate after each step of elimination of short length instances. The ontology is rearranged after each step because there are possibly consecutive instances with the same high-level activity value, after removing some instances.

Table 2.

The calculated probability for one instance, L from LAP, B from BHO, and T from both (predicted and actual target class is "early morning")

| id | L101 | L102 | L103 | L104 | L105 | B101 | B102 | B103 | B104 | B105 | T101 | T102 | T103 | T104 | T105 |
|---|---|---|---|---|---|---|---|---|---|---|---|---|---|---|---|
| 14940 | 0 | 0 | 0 | 0 | 0 | 0 | 0 | 0 | 0.11... | 0.04... | 0 | 0 | 0 | 0.07... | 0.02... |
| 14940 | 0 | 0 | 0.33... | 0 | 0 | 0 | 0 | 0 | 0.11... | 0.04... | 0 | 0 | 0.13... | 0.07... | 0.02... |
| 14940 | 0 | 0 | 0.33... | 0 | 0 | 0 | 0 | 0 | 0.11... | 0.04... | 0 | 0 | 0.13... | 0.07... | 0.02... |
| 14940 | 0 | 0 | 0 | 0 | 0 | 0.13... | 0.09... | 0.35... | 0.07... | 0.13... | 0.09... | 0.04... | 0.17... | 0.04... | 0.06... |
| 14940 | 0 | 0 | 0.33... | 0 | 0 | 0.13... | 0.09... | 0.35... | 0.07... | 0.13... | 0.09... | 0.04... | 0.28... | 0.04... | 0.06... |
| 14940 | 0 | 0 | 0.33... | 0 | 0 | 0.13... | 0.09... | 0.35... | 0.07... | 0.13... | 0.09... | 0.04... | 0.28... | 0.04... | 0.06... |

## 4. Implementation and Experiment

### 4.1. Software for Defining the Assertion Axioms

In the case of user define or edit axioms, a software application is needed. There are two visual interfaces. For BHO, the user is able to define the probability of being in a particular high-level activity while the right hand is in interaction with object X and the left hand is interaction with Y, as shown in Figure 3. There are 23 objects and 529 different states for both hands. However, since many of them are impossible or very unlikely, there is no need to define all of them. For each "hands state", the user sets the probability for all the five high-level activities. The application enforces the constraint that the summation of these five probabilities must be less than or equal to one. If it is less than one, the remaining probability is assigned to idle, which works the same as Null.

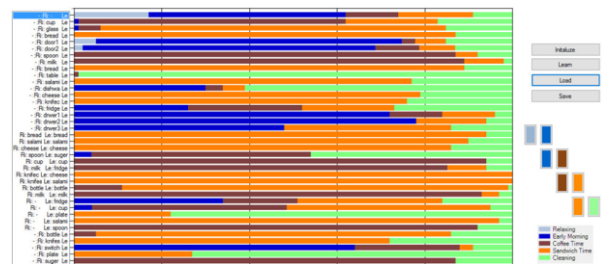

Fig. 3. The software interface for defining Axioms, for BHO.

As shown in Figure 4, the interface is more complicated for LAP; the subject is in a location in the room, their body has an angle toward north, and their posture is sitting, standing, lying, or walking. According to all these parameters, the user determines the probability of being in a particular high-level activity. For this purpose, the user should first choose an activity and then select a particular wedge from a particular circle. A group of circles and angles can be selected as well. The subject can increase or decrease the probability and its color changes accordingly, from white for zero to vivid red for one. The app enforces the probability constraint; the summation of probabilities of each wedge for all high-level activities cannot be more than one. For example, if the probability of a wedge is set to one, vivid red, for coffee time, the probability of the other four activities cannot increase from zero and they remain white if the user attempts to do that.

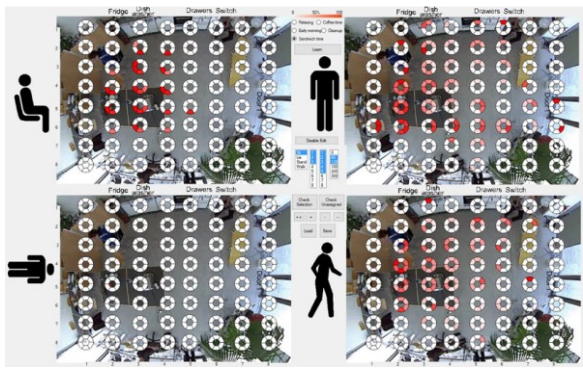

Fig. 4. The software interface for defining Axioms, for LAP.

*4.2. Ontology Storage in a Relational Database*

In this section, we have an overview of storing the ontology on a relational database. The information coming from the processed dataset is stored in the base relations. Each population step of the ontology is performed by running a view on the former version of the ontology. For example, a view receives unsmoothed data from another view and sends the smoothed information to the next view. It should be noted that in modeling, there is no difference between base relations and views; both of them are known as relations or SQL tables [15]. Thus, in any phase, there is a relation that presents the current version of the ontology. In some cases, views are materialized and stored in the form of base relation.

There are automated serial numbers for all tuples in the base SQL tables that are set as the primary key, but are not used in the proposed method. Instead, the probabilistic primary key constraint, as defined in section 3.2, is enforced by defining SQL Server triggers. It guarantees that improper data will not appear in the base relations and, therefore, in the views.

It is unnecessary to include all views and tables here, and only key functions and view should be presented. They are formulated in the relational algebra. It is shorter than an SQL code and easier to understand. The syntax of an SQL code also varies in different DBMSes; for example, a MS SQL Server code usually cannot run by My SQL, for complicated queries, and writing codes from relational algebra formula is easier than converting different SQL codes.

For data smoothing, the following SQL table valued function is defined (it obtains an id number and finds the replacement set of probabilistic values for the location of that instant with interval = 5):

```
datasmooth (@id)  returns @trackingItems (location, pr)
declare @temp (acv,prr) table
declare @sm float
@temp = π arc, pr σ id between (@id - 5) and (@id + 5)
triples_loc
@trackingItems = τ E1 asc π acv, E1 γ acv;
SUM(prr)/11→E1 @temp
@sm = π sm γ ; SUM(pr)→sm @trackingItems
@trackingItems ← π pr trackingItems
Retuen @trackingItems
```

There is a preliminary work on using SQL for classification [8]. For obtaining Bayes axioms, as explained in 3.2.3, these set of consecutive views:

*ranked_obj* query uses smoothing function (nbrfull) to give the smoothed probability of hand-object activity, and its ranking, for each instance. There is a similar query for
$ranked\_obj = \rho\ id←iid, mxpr←pr\ \pi\ iid, hand, pr, rownum()→rnk$ (6)

*mixed_arc* query uses arcfull function to give the smoothed probability and ranking of the mixed angle, posture, and location for each instance.
$mixed\_arc = \rho\ arc←location, mxpr←pr\ \pi\ iid, location, pr, rownum()→rnmx$ (7)

*finalfr_arc* query chooses the most probable arc for each instance. It will be used for training purposes.

*finalfr_arc* = π id, arcc, pr γ id;max(mxpr) → pr,max(arc) → arcc (σrnmx=1 or rnmx=null mixed_loc ⋈ iid = id triples_main)  (8)

*tr_ar* prepares the training data for the angle, posture, and location. The dataset is not annotated for the location and angle. Therefore, the most probable choice is supposed to be true for the training part.
*tr_ar* = π triples_main.id, finalfr_arc.arcc σ ( triples_main.subject = 1 ) or ( triples_main.subject = 2 ) or ( triples_main.subject = 3 ) (finalfr_arc ⋈ finalfr_arc.id = triples_main.id triples_main)  (9)

*c101* gives a number of instances for each BHO. It will be used to discover axioms; a similar query is needed for 102 to 105 classes as well as for LAP.
*c101* = π c101 , codt γ codt;count(tr_ar.id)→c101 (σ Lhlev <> '101' (tr_ar ⋈ tr_ar.id = tr_lab_train.id tr_lab_train))  (10)

*hrul* gives learned axioms for both hands-objects activities.
*hrul* = π tot.codt, c101 / ct→p101, c102 / ct→p102, c103/ ct→p103, c104/ ct→p104, c105 / ct→p105 (tot ⋈ tot.codt = a101.codt a101 ⋈ tot.codt = a102.codt a102 ⋈ tot.codt = a103.codt h103 ⋈ tot.codt = a104.codt a104 ⋈ tot.codt = a105.codt a105)  (11)

At this stage, the assertion axioms, learned or defined by the user, are available. For the reasoning process, the system applies these axioms on instances that should be recognized.

*predict* gives predictions for each instance, separate calculation by the hands and location. Sample results are shown in the left columns of Table 2.
*predict* = (ts_ar ⋈ arcc = arct arul) ⋈ ts_ar.id = tr_lab_test.id ((hrul ⋈ codt = accod ts_ho) ⋈ ts_ho.id = tr_lab_test.id tr_lab_test)  (12)

*mixhoar* presents probabilities in a high-level activity for each instance.
*mixhoar* = π id, Lhlev, 1 - ( 1 - 0.7 * p101 ) * ( 1 - 0.7 * h101 )→v101, 1 - ( 1 - 0.7 * p102 ) * ( 1 - 0.5 *h102 )→v102, 1 - ( 1 - 0.4 *p103 ) * ( 1 - 0.5 *h103)→v103, 1 - ( 1 - 0.9 * p104 ) * ( 1 - 0.7 * h104 )→v104, 1 - ( 1 - 0.6 * p105 ) * ( 1 - 0.5 * h105 )→v105 predict  (13)

## 5. Results and Discussion

### 5.1. Model Evaluation

Zolfaghari et al. [58] proposed six criteria to evaluate human activity recognition models and frameworks and compare them with each other. Here, we exploit these benchmarks to show the proposed system's advantages and drawbacks.

1) Ability to handle uncertain, noisy data and incomplete ambiguous information: The proposed model is fully probabilistic in all stages, which provides a very high ability in dealing with uncertainty, including imperfect data from sensors and non-deterministic axioms.

2) Ability to model complex activity correlations: Unlike OWL-based models, in the proposed model, axioms are not limited to a tree-like schema. A relational model empowers the system to store and manage complex correlations in the form of tuples.

3) Supporting temporal reasoning: In the proposed system, instants are not isolated islands; the value of one instant is related to the values of instances before and after, and the duration of activities is involved in the inference process. If the dataset contains real-time stamps, SQL is able to store and process this type of data.

4) Expressive representation: As the method is not implemented based on knowledge representation languages, the expressive representation benchmark cannot directly apply to it. On the other hand, almost all queries, with more or less effort for query designing, are expressible using SQL. There are only some exceptions like recursive queries that cannot be written in SQL.

5) Reusability: The proposed model is highly reusable; it does not depend on any particular subject, environment, and activity set. In case of changing the location, e.g. from the kitchen to the workhouse, the only necessary change is to redesign the interface of the application of defining LAP axioms.

6) Ability to model complicated activities: The proposed system does not model concurrent, parallel, and multi-subject activities. Opportunity, the applied dataset, does not contain such data and is limited to "daily morning activities" and not to full day activities. Nevertheless, the proposed model is extendable to model more completed activities, as discussed in section 6.

## 5.2. Performance Evaluation

In this section, we evaluate the impact of the proposed and applied methods. The first proposed method is probabilistic data smoothing. It is applied to LAP (location, angle, and posture) and BHO (both hand-object interactions). However, because there is no manual label for the location and angle of the subject, we cannot measure the correctness of the predicted labels for LAP. Labels for the right and left hands interactions with objects are available, and we can compare them with predicted labels, before and after data smoothing.

There are 24 different states for each hand, meaning that there are 576 possible states for both hands. However, there are 479 available states in the dataset. We evaluate the predictions using a simple metric, hit rate: the number of correctly predicted instances to the number of all instances, not only because of numerous classes but also because the first, second, and third probable predictions are important in evaluation. The under/over fill window is equal to 1 second, and the definition is explained later in this section. In about half of instances, both hands are idle, and the value is null. The evaluation is done with and without null values. The results are shown in Table 3.

Table 3.
Hit rate of BHO, before and after the probabilistic data smoothing.

|  | Including nulls | | Excluding nulls | |
| --- | --- | --- | --- | --- |
|  | bf/smooth | af/smooth | bf/smooth | af/smooth |
| 1st prediction | 0.680 | 0.717 | 0.478 | 0.513 |
| 1st or 2nd prediction | 0.781 | 0.807 | 0.627 | 0.658 |
| 1st, 2nd or 3rd prediction | 0.832 | 0.850 | 0.710 | 0.730 |

This system's main goal is to correctly predict high-level activities. The evaluation of the results of this part will reveal the performance of the whole system, including obtaining assertion axioms and the inference process. Since manual labels are in the form of fix length instances, we convert the final ontology to this schema. The assertion axioms are automatically obtained. Some apparent manual modifications in LAP axioms, using the application, lead to some improvement of results. However, because of difficulties in documenting the details of modifications, we ignore the results.

Figure 4 depicts the F-score for predicting high-level activities: (a) considering only the most probable choice or the first reported choice, which means using non-probabilistic input data, (b) considering three most probable choices and their probability values. In both a and b, selection of the target is done with the naïve Bayes method, (c) after three steps of removing short length activities, and (d) after eliminating the under/over fill effect.

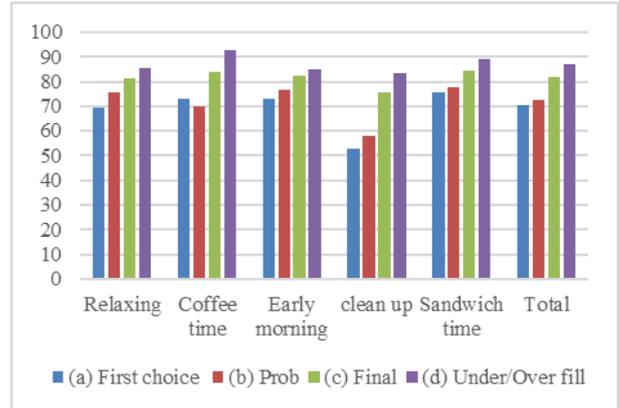

Fig. 4. The F-score for the prediction of high-level activities; (a) using the first choice (most probable) only, (b) using all probable choices (both a and b predictions are made with the help of naïve Bayes), (c) after three steps of removing short activities, named final, and (d) after eliminating the under/over fill effect.

As shown in Figure 1, three most probable choices are more informative, and the proposed method take advantage of this fact to improve results from (a) to (b). After removing short activities which are most likely to be errors, the results are again improved from (b) to (c). As the manual labels are assigned by a human observer, the boundary between activities, e.g. from sandwich time to cleaning, are not very clear, and even different human observers may choose a different time. The automatic system also may determine the starting time of the activity a bit earlier, underfill, or a bit later than what the observer has labeled, overfill. In such a situation, the system performance is underestimated. A window is defined to negate this effect; in (d), if the correct answer is not in the exact place that is predicted, but it is still inside the window, the predicted answer is supposed to be correct. Since the usual length of high-level activities equals several minutes, the window size is 30 seconds. In short, in each step, the obtained results are improved.

Figure 5 compares the actual labels made manually by video checking, with labels made by the system that implements the proposed method before and after three steps of removing short activities, in three consecutive lines. As clearly shown with a magnifier, predicted labels sometimes differ from actual labels, as there are some wrong predictions. The proposed technique of removing short activities in three steps could successfully correct some of these wrong predictions. For instance, the third line of the subject 1 in Figure 5 shows that predicted labels are disturbed with wrong predictions when compared with actual labels. The middle line of the same subject shows that many of these wrong predictions are corrected using the proposed method. What is reported as "Final" in Figure 5 is the same as "Final" in Figure 4, and "Predicted" in Figure 5 is the same as "Prob" in Figure 4. Target wise, the highest improvement more than 17%, goes to cleaning up with this method, which is depicted with green lines in Figure 5.

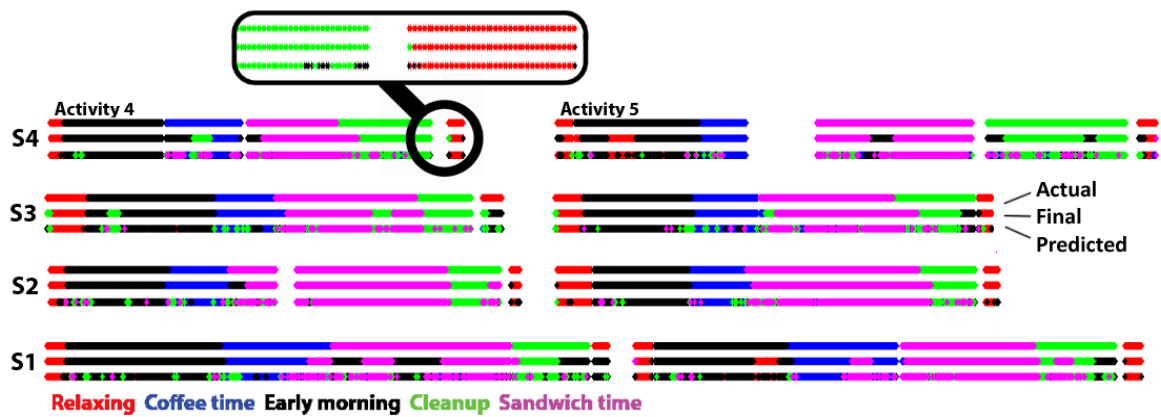

Fig. 5. Comparison of actual, manually labeled activities with predicted labels by the proposed system before and after three steps of removing short activities, for four subjects, each performing two sets of testing activities.

According to [31], KNN has relatively the best performance on the same dataset. We employ KNN on our training and testing sets in the same way that they had used. In addition to the KNN classifier, we compare our approach with that in a previous research applying the same dataset used in the current study. They employed Neutrosophic lattice and fuzzy lattice methods for activity recognition [50]. All these three methods use manual labels of low-level activities provided by the dataset developers. In contrast, the fully probabilistic method, proposed in this research, uses probabilistic predictions computed from sensors' data. In other words, the input data is not error-free in our method, but it is in the other three methods.

As shown in Figure 6, although our proposed method deals with uncertain data from sensors, it outperforms the other methods. For all targets, the fully probabilistic method shows a considerable improvement. Target wise, the highest improvement, compared to the average performance of the other two methods, goes to cleaning up with 30%, while the lowest improvement goes to early morning with 10% improvement. Moreover, the proposed method has less deviation; the performance is almost the same for all targets of high-level activities. The results can be improved even more if we perform probabilistic reasoning on the order of activities. Although in the applied dataset, as shown in Figure 5, all subjects perform activities in the same order.

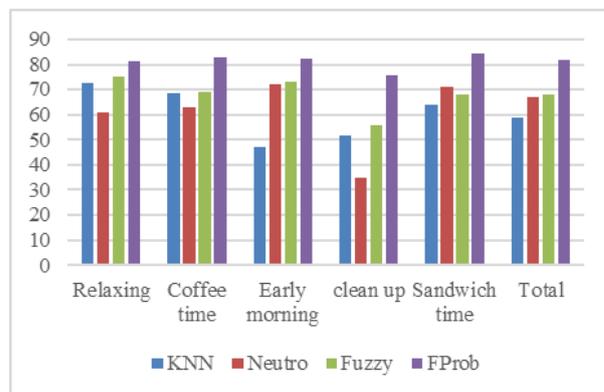

Fig. 6. The performance (F-score) of KNN, Neutrosophic lattice, Fuzzy lattice, and the proposed method: fully probabilistic for five targets of high-level activities.

The running times of queries reported by the SQL Server are recorded. The experiment is carried out on Dell Precision 3620 machine, which is equipped with 3.6GHz Intel i7-4790 quad-core processor and 16 GB RAM running Microsoft SQL Server 2016 express edition, the free version, 64 bits. The training data is about 3 hours and a half, and there is about 2 hours of testing data. With this amount of data, for all queries, including obtaining assertion axioms and the inference process, the running time is reported as zero, meaning that it is less than one second. There are only three exceptions: the running time of procedures for data smoothing of LAP (for all available data, interval size 3) and BHO (for testing data, interval size 5) as well as the three-step elimination of high-level activities reported in Table 4, is longer than the others. In short, the processing time for preprocessing, obtaining axioms, and inference time are very short. However, the applied techniques to improve the system performance are relatively time-consuming. The current system running on the computer mentioned previously for processing the testing data is 27 times faster than a real-time system and processing the training data is 61 times faster.

Table 4.
The running time of procedures in hours, minutes, and seconds, the length of the activity of the processed data, and the ratio of processing to activity time.

| Procedure | Data length | Process time | Ratio |
|---|---|---|---|
| BHO data smooth | 2:02:42 | 0:02:40 | 0.0217 |
| LAP data smooth | 5:30:00 | 0:05:25 | 0.0164 |
| 3-Step proc of HL | 1:49:53 | 0:01:02 | 0.0094 |

## 6. Conclusion and Future Works

This paper presented a set of methods to recognize coarse-grained, high-level human activities from probabilistic fine-grained, low-level activities obtained from location, body, and environmental sensors. The system was modeled in the form of probabilistic ontologies and was implemented in a relational database management system. The method is able to deal with uncertainty, e.g. sensor failure, and it is reusable for different subjects, environments, and activities. The activity recognition performance is relatively high compared to other proposed methods. Moreover, the major advantage of the proposed system is the required performing time, making it highly scalable for batch processing tasks.

The proposed method appears to be capable of being extended to deal with complex high-level activities to discover the order and relationship between them and do temporal reasoning about instances with real timestamps. The selected dataset for this research, Opportunity, had merits and demerits. Near real-world circumstances, fully covered with sensors with some controlled sensor faults and failures in addition to favorable assortment and labeling help us develop a reusable, fully probabilistic ontology-based system for HAR on it. However, it is limited to a short period of daily activities, from waking up in the morning to finishing the breakfast. Even though subjects are free to do each activity in the desired timespan, all high-level activities are performed in the same order. For the purpose of developing more advanced and practical activity recognition systems, superior datasets are required to be recorded with a sufficient sensor coverage in near real-life conditions and with proper labeling, like Opportunity, which contain activities of whole or a major part of the day, are performed on various orders, and include local time. In such a dataset and HAR system, high-level activities are not only obtained from low-level activities, and they can also be obtained from relationships between high-level activities, time, and other available information.

## Acknowledgement

The authors thank Aida Khorshidtalab who kindly read the earlier version of the manuscript and gave us lots of comments and suggestions.